\documentstyle[12pt]{article}

\newcommand{\bb}{\begin{equation}}
\newcommand{\ee}{\end{equation}}
\newcommand{\ba}{\begin{array}}
\newcommand{\ea}{\end{array}}
\newcommand{\beqa}{\begin{eqnarray}}
\newcommand{\eeqa}{\end{eqnarray}}

\newcommand{\MS}{$\bar{MS}$ }  
\newcommand{\lnM}{\ln{\frac{M^2}{\mu^2}}}
\newcommand{\lnm}{\ln{\frac{m^2}{\mu^2}}}

\newcommand{\ep}{effective potential }
\newcommand{\cc}{coupling constant }
\newcommand{\lamb}{\overline{\lambda}}
\newcommand{\lamB}{$\bar{\lambda}$ }
\newcommand{\yyb}{\overline{y}^2}
\newcommand{\yyB}{$\bar{y}^2$ }
\newcommand{\ggsb}{\overline{g}_s^2}

\newcommand{\Mb}{\overline{M}}
\newcommand{\MMb}{\overline{M}^2}
\newcommand{\mb}{\overline{m}}
\newcommand{\mmb}{\overline{m}^2}
\newcommand{\Ib}{\overline{I}}
\newcommand{\zv}{\overline{\zeta}_v}
\newcommand{\rh}{\bar{\rho}}

\newcommand{\gwwb}{\overline{g}_2^2} 
\newcommand{\gwb}{\overline{g}_1^2}

\begin{document}

\title{$\overline{MS}$ Perturbation Theory and the Higgs Boson Mass}
\author{R.S.Willey\thanks{\it Physics Dept, University of Pittsburgh,
Pittsburgh PA 15260;    E-Mail:willey@vms.cis.pitt.edu}} 
 
\date{July 10,1996}

\maketitle 

\begin{center}
Abstract\\
\end{center}

We show that $\overline{MS}$ perturbation theory develops tachyonic 
singularities for some value of the dimensional regularization scale 
$\mu$ unless the physical Higgs mass exceeds some (cutoff dependent) 
minimum value.
\vspace{.3in}

\vspace{.3in}

\clearpage

    For the history of the subject we refer to reviews \cite{Sher} and quote 
only recent papers containing the latest refinements of the conventional 
approach \cite{AI},\cite{ceq}. Once the experimental lower bound on the           top-quark mass 
exceeded $80\,  GeV$, attention shifted from the original Linde-Weinberg bound, 
based on the properties of the one-loop effective potential for small $\phi$, 
close to the minimum, to the large $\phi$ behavior of the \ep , as determined 
by renormalization group (RG) considerations.
\[
V_{eff}(\phi)=\frac{1}{4}\overline{\lambda}(t)(\xi (t)\phi )^4 
\]
Here, $\xi (t)$ is the anomalous dimension factor, and $\overline{\lambda} (t)$ is the 
\MS running \cc . 
\[
t=\ln{\frac{\phi}{M_0}}, \hspace{.5in} \frac{d \lamb (t)}{dt}= 
\beta_{\lambda} (\lamb (t),\overline{g}(t)) 
\]   
It is then argued that vacuum stability requires \lamB $(t) > 0$ up to 
some high scale, $\phi \sim M_{GUT}$,\,or$M_{Pl}$,\,($M_0 \sim M_Z$,\,or $m_t$, or 
$246 GeV$).To implement this condition, one has to know (or approximate) the 
$\beta -$ function, integrate the RG differential equations starting from some 
initial values,\lamB $(0)$,\,$\bar{g}^2(0)$, and relate the smallest acceptable 
\lamB $(0)$ to a physical Higgs mass. (In the latest \ep calculation \cite{ceq} 
 the condition is $\overline{\lambda}_{eff}(t) > 0$ where                         $\bar{\lambda}_{eff}$ differs perturbatively 
from \lamB).

    We propose a new approach. It is also of 
the perturbative RG variety, but is not based on the \ep . 
The results are found to be insensitive to scale ambiguity.
 The essential input is that one is perturbing 
about the correct vacuum. A neccessary condition for this is that the vev of the (shifted) field  be zero, 
order by order in perturbation theory, and the \MS renormalized mass 
squared in the \MS propagator of the shifted field be positive.

   We start by computing the relation between the perturbative pole mass and 
the \MS mass, for both the Higgs boson and the t-quark. The 
relation follows from the perturbative definition of the pole mass, 
\bb
0 =Re\,\overline{D}^{-1}(M^{*^2}) = M^{*^2}-\overline{M}^2                       -Re\,\overline{\Sigma}   (M^{*^2})      \label{defpole} 
\ee 
In this equation, $\bar{D}(q^2)$,\, and $\bar{\Sigma}(q^2)$ are the two-
point Green Function and self-energy function, renormalized according to the 
\MS prescription. $M^*$ is the perturbative pole mass.The result is 
\cite{BW}
\bb
M^{*^2}=\MMb\{1+\lamb (3 \Ib _{00}(M^{*^2})+ 9 \Ib _{\Mb \Mb}(M^{*^2})) + N_c 
\yyb(\frac{M^{*^2}}{\MMb} -4 \frac{\mmb}{\MMb})\Ib_{\mb\mb}(M^{*^2}) + 2( 
\zv -1)\}   \label{MM} 
\ee 
m is the t-quark mass, and y is the t-quark Yukawa coupling ($m = \frac{yv}{    \sqrt{2}}$).    
The contributions from the electroweak gauge sector, proportional to $g_2,g_1$, 
have also been calculated, but are not written out here.
they will be included below. The term $\zv -1$ comes from a finite shift 
of the vev required in the \MS scheme to enforce $<\hat{H}>=0$ through one- 
loop order. It will cancel out of the ratio computed below, so we do not have 
to give its value here. \cite{BW}$ \bar{I}_{ab}$ is the dimensionally  
regularized \MS scalar one-loop two-point integral. 
\bb
 \ba{c}
  \Ib_{ab}(q^2)=[\mu^{4-d} i\int \frac{d^d l}{(2\pi)^d}\frac{1}{(l^2 -a^2) 
   ((l-q)^2 -b^2)}]_{_{\overline{MS}}}  \\    
  = \frac{1}{16\pi^2}[\ln{\frac{ab}{\mu^2}} + \int_{0}^{1} dx \ln{\frac{a^2 x + 
  b^2 (1-x) -q^2 x(1-x)}{ab}}]   
 \ea 
 \label{Iab}   
\ee
Then (\ref{MM}) is
\bb
 M^{*^2}=\MMb\{1 + \frac{\lamb}{16 \pi^2}[12 \lnM -24 + 3\sqrt{3}\pi] + N_c 
 \frac{\yyb}{16 \pi^2}(1-\frac{4}{r^2})[\lnm + f(r)] + 2(\zv -1)\}  
  \label{MM2} 
\ee 
where  
\[ 
 r=\frac{M}{m}, \hspace{.5in} f(r)=-2+2\sqrt{\frac{4-r^2}{r^2}}\arctan{
 \sqrt{\frac{r^2}{4-r^2}}} 
\]
The corresponding calculation for the t-quark gives  \cite{BW1}
\bb
 m^{*^2}=\mmb \{1+\frac{\yyb}{16 \pi^2}[\frac{3}{2}\lnm +\Delta (r)] +
 \frac{\ggsb}{16 \pi^2}C_F(8-6\lnm) + 2(\zv -1)\}  \label{mm} 
\ee
where
\[
 \Delta(r)= -4+\frac{r^2}{2}+(\frac{3}{2}r^2 -\frac{1}{4}r^4)\ln{r^2}+\frac
 {r}{2}(4-r^2)^{\frac{3}{2}}\arctan{\sqrt{\frac{4-r^2}{r^2}}} 
\]
We take the ratio of (\ref{MM}) to (\ref{mm}) and expand to one-loop order. 
\bb
 \ba{c}
  \frac{M^{*^2}}{m^{*^2}}=\frac{\MMb}{\mmb}\{1+\frac{\lamb}{16 \pi^2}[12\lnM - 
  24+3\sqrt{3}\pi] + \frac{\yyb}{16 \pi^2}[N_c(1-\frac{4}{r^2})(\lnm +f(r))- 
  \frac{3}{2}\lnm -\Delta (r)] \\                                                  +\frac{\ggsb}{16 \pi^2}C_F(6\lnm -8) + g_2^2,g_1^2 \, terms + 2-loop\} 
 \ea
  \label{rat1}  
\ee 
The $\zv -1$ terms, which also contain explicit dependence on $\ln{\mu^2}$, 
have cancelled out.
A necessary condition for the \MS perturbation calculations to be defined 
in the broken symmetry phase is that $\bar{M}^2,\bar{m}^2$ be positive.
Since the ratio of pole masses is positive, (\ref{rat1}) satisfies the 
requirement perturbatively, for $\mu$ around the weak scale. For large 
$\mu^2$, one has to provide a RG treatment of the large logarithms, just 
as in the conventional calculation involving the \ep.
In the broken symmetry phase, one can define the renormalized coupling 
constants such that the relation 
\bb
 \frac{M^2}{m^2}=4\frac{\lambda}{y^2}  \label{ccrat} 
\ee
is exact when all the quantities are either ''star'' (on-shell renormalization 
scheme) or ''bar''(\MS renormalization scheme)\cite{BW1},\cite{BW}. Thus, not   all    
quantities in (\ref{rat1}) can be varied independently as functions of $\mu$. 
we use (\ref{ccrat}) to eliminate \lamB appearing in (\ref{rat1}). 
 To leading (one-loop) order, the scale dependence of  
the ratio of \MS masses is determined by the coefficients of the explicit 
$\ln{\mu^2}$ terms in (\ref{rat1}). For the other masses in (\ref{rat1}), the 
difference between ''star'' and ''bar'' is higher order (combined with 
explicitly two-loop effects),as is the implicit $\mu$ dependence of the 
''bar'' coupling constants. After these observations, and reinstating the 
$g_2^2,g_1^2$ terms, differentiating (\ref{rat1}), we obtain
\bb
 \ba{c} 
  \mu \frac{d}{d\mu}(\rh)=\frac{\yyb}{16 \pi^2}[6 \rh ^2 +(2N_c -3 +12 C_F 
  \frac{\ggsb}{\yyb})\rh -8 N_c  \\ 
  -(\frac{9}{2}\frac{\gwwb}{\yyb}+\frac{1}{6}\frac{\gwb}{\yyb})\rh 
  +3\frac{\overline{g}_2^4}{\overline{y}^4} 
  +\frac{3}{2}(\frac{\gwwb +\gwb}{\yyb})^2] + 2-loop 
 \ea 
  \label{deriv} 
\ee
 
where $\rh = r^2 = \frac{\bar{M}^2}{\bar{m}^2}$. 

  Let the right hand side of (\ref{deriv}) be denoted $\beta_{\rho}$.  
Because of the $- 8 N_c$ term in (\ref{deriv}), there is a critical  
value of $\rho$ below which $\beta_{\rho}$ becomes negative. And if the 
starting value of $\rho$ is below this value, as $\rho$ decreases the 
derivative becomes more negative, driving $\rho$ negatuve for some value 
of $\mu$, unless some higher order effect intervenes. 

  The first higher order effect is the running of the \MS coupling constants, 
which appear as coefficients in (\ref{deriv}), and the dependence of the 
lower bound on the cutoff (maximum value of $\frac{\mu}{\mu_0}$). One has to  
integrate the coupled RG equations for five independent "coupling constants", 
$\bar{g}_s^2, \bar{g}_2^2, \bar{g}_1^2, \bar{y}^2, \bar{\rho}$. Let $t =  
\ln{\frac{\mu}{\mu_0}}$.  

\bb
 \ba{c}
  \frac{d}{dt}\ggsb = -\frac{1}{16 \pi^2}(22-\frac{4}{3}N_f)\overline{g}_s^4
      \\
    \frac{d}{dt}\overline{g}_2^2 = -\frac{1}{16 \pi^2}[\frac{44}{3}-\frac{4}
     {3}N_f -\frac{1}{3}N_d]\overline{g}_2^4  \\
    \frac{d}{dt}\overline{g}_1^2 = \frac{1}{16 \pi^2}[\frac{20}{9}N_f +\frac
     {1}{3}N_d]\overline{g}_1^4  \\
    \frac{d}{dt}\yyb = \frac{1}{16 \pi^2}[(3+2 N_c)\overline{y}^4
     -12 C_F \ggsb \yyb -\frac{9}{2}\overline{g}_2^2\yyb -\frac{17}{6}
    \overline{g}_1^2\yyb]  \\
    \frac{d}{dt}(\rho)=\frac{\yyb}{16 \pi^2}[6 \rho^2 +(2N_c -3 +12 C_F
  \frac{\ggsb}{\yyb})\rho -8 N_c  \\
  -(\frac{9}{2}\frac{\gwwb}{\yyb}+\frac{1}{6}\frac{\gwb}{\yyb})\rho
  +3\frac{\overline{g}_2^4}{\overline{y}^4}
  +\frac{3}{2}(\frac{\gwwb +\gwb}{\yyb})^2]
 \ea
  \label{5diff}
\ee

The first three equations are integrated trivially. If we neglect the $\bar
{g}_2,\bar{g}_1$ contributions to the $\bar{y}$ running, that equation can 
also be integrated analytically. But if one runs up to high scales, the 
electroweak gauge couplings become of same order as the QCD coupling constant; 
so we use NDSolve from Mathematica to provide an interpolating function 
solution for \yyB which is substituted into the $\bar{\rho}$ equation, which is  again 
integrated numerically by NDSolve. 

  Before giving any results, we discuss the question of their sensitvity to 
the choice of dimensional regularization scale $\mu$. In this approach, the 
scale sensitvity comes from the choice of the Electroweak scale $\mu_0$. 
This enters in three ways: (i) the starting values of the $\bar{g}_i^2, 
\bar{y}^2$ at $t = 0$ ($\mu = \mu_0$).  (ii) the connection between $t_{max}$ 
and the nominal value of the cutoff,  $\mu_{max} = \Lambda$.  (iii) the 
conversion of the found critical $\rh_c (0)$ back to the ratio of pole 
masses, (\ref{rat1}). It is clear that there will be substantial cancellation 
between these.  

  We take $m_t = 175$ GeV. For the low scale cutoff, we take $\Lambda = 
1$ TeV. For a first orientation, we keep only the large coupling constants       , $\, g_s,y$, 
in equations (\ref{5diff}), ($g_2,g_1 \rightarrow 0$). To check the 
sensitivity to the choice of starting Electroweak scale $\mu_0$, we 
solve the remaining three equations from (\ref{5diff}) starting from 
$\mu_0 = M_Z$ ($t_{max} = 2.4$), and again, starting from $\mu_0 = m_t$  
($t_{max} = 1.74$). We also require as input the initial values   
$\bar{g}_s^2,\bar{y}^2$. For $mu_0 = M_Z$, We take                               $\bar{g}_s^2 = 1.483 \;(\alpha_s(M_Z)= .118$).  
For the Yukawa coupling constant we have  
$$ v^*(M_W) = (\sqrt{2} G_F)^{-\frac{1}{2}}\;(1-\Delta r^*)^{-\frac{1}{2}} 
= 251 \mbox{GeV}  $$  \\
$$ y^{*^2}(M_Z) = 2 \frac{m^{*^2}}{v^{*^2}} = .972 $$
Then we use equation (59) of \cite{BW} to convert from star to bar, giving 
 the initial value $\bar{y}^2(M_Z) = .970$. With these initial values, the 
critical initial value of $\rho(0) $ for which $\rho(t)$ falls through zero 
at $t = 2.4$ is $\rho_c(0) = .26685 = (\bar{M}/\bar{m})_c^2 $. Converting  
back to a ratio of pole masses by (\ref{rat1}) gives the critical value 
$M^*_c = 74$ GeV.

 For initial scale $\mu_0 = m_t$, we need $\bar{g}_s^2(m_ta)$ 
and $\bar{y}^2(m_t)$. For the QCD \cc , to our level of accuracy, we can 
simply run $\bar{g}_s^2$ from $\mu = M_Z$ to $\mu = m_t$ using the first 
equation of (\ref{5diff}). This gives $\bar{g}_s^2(m_t) = 1.355 \;$              $(\alpha_s(m_t)  = .1078)$. 
The determination of the Yukawa \cc is more complicated. We can proceed in 
two different orders. First, we can just run $\bar{y}^2(M_Z)$ to 
$\bar{y}^2(m_t)$ using the fourth equation of (\ref{5diff}) and obtain 
$\bar{y}^2(m_t) =  .905$.
Alternatively, we can run $v^{*^2}(M_Z)$ to $v^{*^2}(m_t)$ (see equation 
(40) of (\cite{BW})) and convert to $y^{*^2}(m_t)$ and then to                  $\bar{y}^2(m_t)$ which gives $\bar{y}^2(m_t) = .877.$    
 So we carry through the calculation for three different initial values:  
$\bar{y}^2(m_t) = .905,.891 (avg) , .877$. The resulting $M^*_C$ are 
(resp) $74.7, 74.1, 73.5$. These results give us some confidence that the  
calculation is not sensitive to the choice of initial weak scale in the 
range from $M_Z$ to $m_t$, as long as the scale dependence of the input 
parameters is handled consistently.   

    Having checked this point, we reinstate the gauge coupling constants 
in(\ref{5diff}). We integrate this set of equations, starting at $\mu_0 = m_t$ 
, with the additional input $g_2^2(m_t) = .4239,\; g_1^2(m_t) = .1260$. Then the
 critical initial value of $\rho(0)$ for which $\rho(t)$ falls through zero at 
$t = 1.74$ is $.182$. Converting back to a ratio of pole masses by equation 
(\ref{rat1}) gives $M^*_c = 72$ GeV. We take this value as our best estimate 
of the smallest Higgs mass for which the $\overline{MS}$ perturbation theory 
is nontachyonic up to scale $\mu$ equal to  one TeV (for $m_t = 175$ GeV). 

 Taking the large scale to be the Planck scale ($\mu_{max}\simeq 10^{19}, 
t_{max}\simeq 39 $) and using the same input at $\mu_0 = m_t$, we find the 
critical value of the Higgs mass to be $140$ GeV. 

   In conclusion, we find that the $\overline{MS}$ perturbation theory 
develops tachyonic singularities (negative mass squared in the $\overline{MS}$  renormalized 
propagator) when the dimensional regularization scale factor $\mu$ exceeds 
some assigned cutoff value unless the physical Higgs mass exceeds some 
minimum value, which depends on the cutoff value. 

  The requirement that $\overline{MS}$ perturbation theory not develop 
tachyonic singularites below some prescribed cutoff scale is in principle 
different from and independent of the requirement that the effective potential 
not develop a stable minimum at some value of the vev much greater than the 
weak scale. However,
 until the recent papers of
Casas,Espinosa, and Quiros \cite{ceq} it has been generally taken that the       requirement
on the effective potential was practically equiivalent to the requirement
that the $\overline{MS}$ running quartic coupling constant stay positve
below the cutoff scale. And in the minimal Standard Electroweak Model, in
conventional renormalization schemes, the condition of positive ratio of
squared masses implies, and is implied by, the condition of positive quartic and
squared Yukawa coupling constants (\ref{ccrat}). In this case the numerical       results are
not different (up to differences in handling input parameters,etc); and 
our numerical results are the same as those of Altarelli and Isadori 
\cite{AI}. 
 But
CEQ have argued that consideration of the scale dependence involved in the
minimization of the perturbative effective potential leads to the
requirement that$\overline{\lambda}_{eff}$ stay positive, where $\overline{
\lambda}_{eff}$ is not the same as $\overline{\lambda}$. Then the numerical
equivalence is broken and the results of the two approaches are different.
It is also possible that the singular behavior we have found is just a 
pathology of $\overline{MS}$ perturbation theory, which is interesting in 
itself, but not justifcation for a restriction on the physical Higgs mass.

It is clearly desirable to have a large scale lattice simulation study of the 
combined Higgs-heavy quark-QCD sector. (Contributions from light quarks and 
electroweak gauge bosons are small, particularly if one doesn't run up to some  very high scale).  We note that a quenched approximation 
simulation is not adequate for this problem. The term in (\ref{deriv}) 
which triggers the possible instability is the $-8 N_c$, clearly a contribution  from an internal fermion closed loop.

\end{document}